\documentclass[prl,twocolumn]{revtex4}%
\usepackage{amssymb}
\usepackage{graphicx}
\usepackage{dcolumn}
\usepackage{bm}
\usepackage{amsmath}
\usepackage{amsfonts}%
\begin{document}
\title{Bell-Correlated Activable Bound Entanglement in Multiqubit Systems}
\author{Somshubhro Bandyopadhyay}
\email{sbandyop@chem.utoronto.ca, som@ee.ucla.edu}
\affiliation{Centre for Quantum Information and Quantum Control and Department of
Chemistry, University of Toronto, Toronto, ON M5S 3H6, Canada. \ \ \ \ \ \ \ \ \ \ \ \ \ \ \ \ \ \ }
\author{Indrani Chattopadhyay}
\email{ichattopadhyay@yahoo.co.in}
\affiliation{Applied Mathematics Department, University of Calcutta, Kolkata, India.}
\author{Vwani Roychowdhury}
\email{vwani@ee.ucla.edu}
\affiliation{Electrical Engineering Department, UCLA, Los Angeles, CA 90095, USA.}
\author{Debasis Sarkar}
\email{debasis@cubmb.ernet.in, debasis1x@yahoo.co.in}
\affiliation{Applied Mathematics Department, University of Calcutta, Kolkata, India.}

\begin{abstract}
{\small We show that the Hilbert space of even number ($\geq4$) of qubits can
always be decomposed as a direct sum of four orthogonal subspaces such that
the normalized projectors onto the subspaces are activable bound entangled
(ABE) states. These states also show a surprising recursive relation in the
sense that the states belonging to $2N+2$ qubits are Bell correlated to the
states of $2N$ qubits; hence, we refer to these states as Bell-Correlated ABE
(BCABE) states. We also study the properties of noisy BCABE states and show
that they are very similar to that of two qubit Bell-diagonal states.}

\end{abstract}
\maketitle



\section{Introduction and Results}

The quantum states that are not distillable \cite{distill} under local
operations and classical communications (LOCC) despite being inseparable are
said to be bound entangled (BE) \cite{horo97, horoetal97, peresbruss,
horoetal98, upb, wernerwolf00}. Bound entangled states exhibit a new kind of
irreversibility in physics where one has to spend finite amount of
entanglement to prepare such states but one cannot extract any non-zero amount
of entanglement from such states via LOCC. Thus the amount of entanglement of
formation is irreversibly lost during the state preparation. Recent studies
involving bound entangled states include characterization of such states
\cite{BGR04, terhalhoro00, sanpera00, horolewenstein00, HSTT}, violation of
Bell type inequalities \cite{Dur01, Acin02, Dagomir02, Horo04} and possible
practical applications \cite{horoopenheim03, muraovedral00}.

For bipartite systems, bound entanglement is clearly defined as it involves
only two spatially separated parties and a necessary and sufficient condition
for distillability of bipartite quantum states is known \cite{horoetal98}. In
a multiparty setting, however, due to several distinct spatially separated
configurations, the definition of bound entanglement is not unique. A
multipartite quantum state is said to be bound entangled if there is no
distillable entanglement between any subset as long as \textit{all} the
parties remain spatially separated from each other. When, however, one also
allows some of the parties to group together and perform local operations
collectively, two qualitatively different classes of bound entanglement arise:
(a) Activable Bound Entangled (ABE) states -the states that are not
distillable when every party is separated from every other but becomes
distillable, if certain parties decide to group together \cite{Dur00,smolin01}%
. This implies that there is at least one bipartite partition/cut where the
state is negative under partial transposition (NPT) \cite{peres95horo96}. Such
states have been also referred to as \emph{Unlockable Bound Entangled} (UBE)
states in the literature. (b) \emph{Non-Activable Bound Entangled} states
-states that are not distillable under any modified configuration as long as
there are at least two spatially separated groups. In other words, such states
are always positive under partial transposition across any bipartite partition
\cite{upb}.

Despite recent studies, the distribution and structure of such states in the
Hilbert space have not been explicitly studied. In this work we show that
bound entangled states have natural existence in the structure of the Hilbert
space of even number, $2N+2$, of qubits(when $N\geq1$). In particular, the
Hilbert space of $2N+2$ qubits, $N\geq1,$ can be decomposed as a direct sum of
four orthogonal subspaces such that the normalized projector onto each
subspace is an activable bound entangled state. The set of four ABE states are
shown to be unitarily related to each other via a local pauli operator on one
of the qubits. Surprisingly, the states exhibit a recursive property, i.e.,
each state of $2N+2$ qubits can be expressed as a convex combination (with
equal weights) of four two-qubit Bell states correlated with the four ABE
states of $2N$ qubits. The only exception occurs for four qubit states, where
the Bell states of two qubits are correlated to Bell states of the other two
qubits. It is interesting to note that one of these four ABE states for the
four-qubit system has been previously discovered by Smolin \cite{smolin01}. We
call these $(2N+2)$-qubit ABE states Bell-correlated activable bound entangled
states (BCABE). 

As noted before the bound entangled states that we present in this work are
activable. In such an activable configuration we find that the distillable
entanglement between any two parties is always one ebit and therefore
independent of $N.$ We also study properties of the noisy BCABE states. The
noisy states are constructed by taking a convex combination of the four BCABE
states. Remarkably, the entanglement properties of these noisy $2N+2$ qubit
bound entangled states can be directly mapped onto that of two qubit Bell
diagonal states.

\section{Hilbert space of 2N+2 qubits: Decomposition and bound entangled
states}

Consider now a system of $2N$ qubits. Let $\left\vert p_{i}\right\rangle
=\left\vert a_{1}^{i}a_{2}^{i}...a_{2N}^{i}\right\rangle $ where $a_{1}
^{i}=0$, and $a_{j}^{i}\in\{0,1\}$, for all $j=2, \cdots,2N$ such that there
is an even number of 0s in the string $a_{1}^{i}a_{2}^{i}...a_{2N}^{i}.$
Likewise, let $\left\vert q_{i}\right\rangle =\left\vert b_{1}^{i}b_{2}%
^{i}...b_{2N}^{i}\right\rangle ,$ where $b_{1}^{i}=0,$ and $b_{2}%
^{i},...,b_{2N}^{i}$ are either 0 or 1 with odd number of 0s in the string
$b_{1}^{i}b_{2}^{i}...b_{2N}^{i}$. One can also define the states orthogonal
to $\left\vert p_{i}\right\rangle ,\left\vert q_{i}\right\rangle $ as:
$\left\vert \overline{p_{i}}\right\rangle =\left\vert \overline{a_{1}^{i}%
}\overline{a_{2}^{i}}...\overline{a_{2N}^{i}}\right\rangle $ and $\left\vert
\overline{q_{i}}\right\rangle =\left\vert \overline{b_{1}^{i}}\overline
{b_{2}^{i}}...\overline{b_{2N}^{i}}\right\rangle $ where $\left\langle
\overline{a_{j}^{i}}|a_{j}^{i}\right\rangle =0 =$ $\left\langle \overline
{b_{j}^{i}}|b_{j}^{i}\right\rangle ,\forall j=1,...,2N$ and $i=1,...,2^{2N-2}%
.$ Note that the four sets of states, defined by $\left\vert p_{i}%
\right\rangle $'s, $\left\vert \overline{p_{i}}\right\rangle $'s, $\left\vert
q_{i}\right\rangle $, and $\left\vert \overline{q_{i}}\right\rangle $'s
respectively, are non-overlapping and all have same cardinality, and they
together span the complete Hilbert space of $2N+2$ qubit systems.

Now we define the following four sets of states:%
\begin{align}
S_{\Phi}^{\pm}  &  =\left\{  \left\vert \Phi_{i}^{\pm}\right\rangle =\frac
{1}{\sqrt{2}}\left(  \left\vert p_{i}\right\rangle \pm\left\vert
\overline{p_{i}}\right\rangle \right)  ,i=1,...,2^{2N-2}\right\}  \label{set1}%
\end{align}
\begin{align}
S_{\Psi}^{\pm}  &  =\left\{  \left\vert \Psi_{i}^{\pm}\right\rangle =\frac
{1}{\sqrt{2}}\left(  \left\vert q_{i}\right\rangle \pm\left\vert
\overline{q_{i}}\right\rangle \right)  ,i=1,...,2^{2N-2}\right\}  \label{set2}%
\end{align}

We can associate with every set $S,$ a subspace of the complete Hilbert space
where the states belonging to $S$ span that subspace and all the subspaces are
orthogonal to each other. In terms of Hilbert space decomposition we can write
this as%
\begin{equation}
H=H_{\Phi}^{+}\oplus H_{\Phi}^{-}\oplus H_{\Psi}^{+}\oplus H_{\Psi}^{-}%
\end{equation}

Observe that together the states span the full Hilbert space and often this
basis is referred to as the cat or GHZ basis.

We will use the notation $\left[  \cdot\right]  $ for pure state projector
$\left\vert \cdot\right\rangle \left\langle \cdot\right\vert $. Let us now
define the unnormalized projectors on to the subspaces spanned by the set of
states given by Eqs. (\ref{set1},\ref{set2}):
\begin{equation}
P_{2N}^{\pm}=\sum_{i=1}^{2^{2N-2}}\left[  \Phi_{i}^{\pm}\right]  ; Q_{2N}%
^{\pm}=\sum_{i=1}^{2^{2N-2}}\left[  \Psi_{i}^{\pm}\right]
\end{equation}

The set of above four projectors are connected to each other by one pauli
operation on one qubit. For instance, consider the unitary operators $U_{i}=
I_{1}\otimes..\otimes I_{2N-1}\otimes\sigma_{2N}^{i}$, where $i\in\{z,x,y\}$,
i.e., $U_{i}$ applies the $i^{th}$ Pauli operator on the $(2N)^{th}$ qubit.
Then one can verify that%

\begin{align}
P_{2N}^{-}  &  =U_{z} P_{2N}^{+} U_{z}^{\dag}\ ,\\
Q_{2N}^{+}  &  =U_{x} P_{2N}^{+} U_{x}^{\dag}\ ,\\
Q_{2N}^{-}  &  =U_{y} P_{2N}^{+} U_{y}^{\dag}\ .
\end{align}

We will now show how to generate the above set of four projectors in the case
$2N+2$ qubits starting from the set of \textit{2N} qubits. First one can write
$P_{2N+2}^{+}$ as%
\begin{equation}
P_{2N+2}^{+}=\sum_{i=1}^{2^{2N-2}}%
{\textstyle\sum\limits_{k=1}^{k=4}}
\left[  \Omega_{i}^{k}\right]
\end{equation}

where the $\Omega$ states are defined as:%
\begin{align}
\left\vert \Omega_{i}^{1}\right\rangle  &  =\frac{1}{\sqrt{2}}\left(
\left\vert 00\right\rangle \left\vert p_{i}\right\rangle +\left\vert
11\right\rangle \left\vert \overline{p_{i}}\right\rangle \right) \nonumber\\
\left\vert \Omega_{i}^{2}\right\rangle  &  =\frac{1}{\sqrt{2}}\left(
\left\vert 11\right\rangle \left\vert p_{i}\right\rangle +\left\vert
00\right\rangle \left\vert \overline{p_{i}}\right\rangle \right) \nonumber\\
\left\vert \Omega_{i}^{3}\right\rangle  &  =\frac{1}{\sqrt{2}}\left(
\left\vert 01\right\rangle \left\vert q_{i}\right\rangle +\left\vert
10\right\rangle \left\vert \overline{q_{i}}\right\rangle \right) \nonumber\\
\left\vert \Omega_{i}^{4}\right\rangle  &  =\frac{1}{\sqrt{2}}\left(
\left\vert 10\right\rangle \left\vert q_{i}\right\rangle +\left\vert
01\right\rangle \left\vert \overline{q_{i}}\right\rangle \right)
\end{align}

Now recall that,
\begin{align}
\left\vert 00\right\rangle  &  =\frac{1}{\sqrt{2}}\left(  \left\vert \Phi
^{+}\right\rangle +\left\vert \Phi^{-}\right\rangle \right)  ,\left\vert
11\right\rangle =\frac{1}{\sqrt{2}}\left(  \left\vert \Phi^{+}\right\rangle
-\left\vert \Phi^{-}\right\rangle \right)  ,\nonumber\\
\left\vert 01\right\rangle  &  =\frac{1}{\sqrt{2}}\left(  \left\vert \Psi
^{+}\right\rangle +\left\vert \Psi^{-}\right\rangle \right)  ,\left\vert
10\right\rangle =\frac{1}{\sqrt{2}}\left(  \left\vert \Psi^{+}\right\rangle
-\left\vert \Psi^{-}\right\rangle \right)  \label{prodtobell}%
\end{align}

where the two qubit Bell states are defined by
\begin{equation}
\left\vert \Phi^{\pm}\right\rangle =\frac{1}{\sqrt{2}}\left(  \left\vert
00\right\rangle \pm\left\vert 11\right\rangle \right)  ,\left\vert \Psi^{\pm
}\right\rangle =\frac{1}{\sqrt{2}}\left(  \left\vert 01\right\rangle
\pm\left\vert 10\right\rangle \right)
\end{equation}

Substituting the above in the expression for $\Omega,$ and after some
algebraic manipulations, one obtains%
\begin{align}
P_{2N+2}^{+} &  =\left[  \Phi^{+}\right]  \otimes P_{2N}^{+}+\left[  \Phi
^{-}\right]  \otimes P_{2N}^{-}+\left[  \Psi^{+}\right]  \otimes Q_{2N}%
^{+}\nonumber\\
&  +\left[  \Psi^{-}\right]  \otimes Q_{2N}^{-}%
\end{align}

This recursive form is particularly illuminating. However at this point let us
normalize the above projector to make it a legitimate density matrix and write
it as:%
\begin{align}
\rho_{2N+2}^{+}  &  =\frac{1}{4}(\left[  \Phi^{+}\right]  \otimes\rho_{2N}%
^{+}+\left[  \Phi^{-}\right]  \otimes\rho_{2N}^{-}+\left[  \Psi^{+}\right]
\otimes\sigma_{2N}^{+}\nonumber\\
&  +\left[  \Psi^{-}\right]  \otimes\sigma_{2N}^{-}) \label{rhoplus}%
\end{align}
where%
\begin{equation}
\rho_{2N}^{\pm}=\frac{1}{2^{2N-2}}P_{2N}^{\pm},\sigma_{2N}^{\pm}=\frac
{1}{2^{2N-2}}Q_{2N}^{\pm}%
\end{equation}

Let us now look at the properties of the state $\rho_{2N+2}^{+}$ more closely.

\begin{itemize}
\item By construction, the state is invariant under interchange of parties. To
see this, consider the unnormalized state $P_{2N}^{+}.$ This projector is an
equally weighted convex combination of the states belonging to the set
$S_{\Phi}^{+}.$ Now if we interchange the qubits, then under any such
permutation the states belonging to the set just map onto each other leaving
the whole projector invariant. In other words if one denotes the $j^{th}$
party as $A_{j},$ then $\rho\left(  \cdots,A_{i},\cdots,A_{k},\cdots\right)
=\rho\left(  \cdots,A_{k},\cdots,A_{i},\cdots\right)  $ for all possible
$i,k.$

\item The state is entangled. One way to see is that if any 2N parties come
together and do a joint measurement to discriminate the states $\left\{
\rho_{2N}^{+}, \rho_{2N}^{-},\sigma_{2N}^{+},\sigma_{2N}^{-}\right\}  $ (as
they are mutually orthogonal), then this will result in a maximally entangled
state between the remaining two. Or else,\textit{ N} parties could pair up and
do sequential Bell measurements on their two qubits which will lead to
distillation of a maximally entangled state between the remaining two who
didn't come together and remained spatially separated. Therefore the state
must be entangled to begin with, otherwise no configuration could allow any
entanglement to be distilled between separated parties.

\item When all the \textit{2N+2} parties remain spatially separated, then the
state is not distillable as it is separable across every \textit{2 : 2N}
bipartite cut. This is easily seen as the state itself is written in a
\textit{2 : 2N} separable form. That it is separable across every such cut
follows from the permutation symmetry. This makes every party separated from
every other by at least one separable cut and hence no entanglement can be distilled.
\end{itemize}

As the state is entangled but not distillable when \textit{all} the parties
are separated from each other the state must be a bound entangled state. Since
the state becomes distillable if a subset of the parties come together and
perform collective LOCC, the state is activable. Hence the state is an ABE state.

For \textit{2N+2} qubits one can generate the other three states following the
same prescription. However it is much simpler by noting that the states are
all single Pauli connected. Explicitly the remaining three states can be
written using Eqs.(5, 6, 7) as:%
\begin{align}
\rho_{2N+2}^{-} &  =\frac{1}{4}(\left[  \Phi^{+}\right]  \otimes\rho_{2N}%
^{-}+\left[  \Phi^{-}\right]  \otimes\rho_{2N}^{+}+\left[  \Psi^{+}\right]
\otimes\sigma_{2N}^{-}\nonumber\\
&  +\left[  \Psi^{-}\right]  \otimes\sigma_{2N}^{+})\label{rhominus}%
\end{align}
and
\begin{align}
\sigma_{2N+2}^{\pm} &  =\frac{1}{4}(\left[  \Psi^{+}\right]  \otimes\rho
_{2N}^{\pm}+\left[  \Psi^{-}\right]  \otimes\rho_{2N}^{\mp}+\left[  \Phi
^{+}\right]  \otimes\sigma_{2N}^{\pm}\nonumber\\
&  +\left[  \Phi^{-}\right]  \otimes\sigma_{2N}^{\mp})\label{sigmaplusminus}%
\end{align}

The above results can now be summarized in the form of a theorem.

\textbf{Theorem }T\textit{he Hilbert space of }$2N+2$\textit{ qubits, }%
$N\geq1,$ \textit{ can always be decomposed as a direct sum of four orthogonal
subspaces such that the normalized projectors onto the subspaces are activable
bound entangled states.}

Let us note here that when $N=1,$ the set of states $\left\{  \rho_{4}^{\pm
},\sigma_{4}^{\pm}\right\}  $ are Bell correlated to the set of states
$\left\{  \rho_{2}^{\pm},\sigma_{2}^{\pm}\right\}  $, which are not bound
entangled but maximally entangled and hence distillable. However this case is
the only exception when the set of bound entangled states $\left\{
\rho_{2N+2}^{\pm},\sigma_{2N+2}^{\pm}\right\}  $ is not Bell correlated to the
set of bound entangled states $\left\{  \rho_{2N}^{\pm},\sigma_{2N}^{\pm
}\right\}  $.

\section{Illustrations}

\subsection{The Hilbert Space of Four Qubits}

Consider the four sets of states as defined before in Eq (\ref{set1}%
,\ref{set2}).
\begin{align}
S_{\Phi}^{\pm}  &  =\left\{
\begin{array}
[c]{c}%
\frac{1}{\sqrt{2}}\left(  \left\vert 0000\right\rangle \pm\left\vert
1111\right\rangle \right)  ,\frac{1}{\sqrt{2}}\left(  \left\vert
0011\right\rangle \pm\left\vert 1100\right\rangle \right)  ,\\
\frac{1}{\sqrt{2}}\left(  \left\vert 0101\right\rangle \pm\left\vert
1010\right\rangle \right)  ,\frac{1}{\sqrt{2}}\left(  \left\vert
0110\right\rangle \pm\left\vert 1001\right\rangle \right)
\end{array}
\right\} \nonumber\\
S_{\Psi}^{\pm}  &  =\left\{
\begin{array}
[c]{c}%
\frac{1}{\sqrt{2}}\left(  \left\vert 0001\right\rangle \pm\left\vert
1110\right\rangle \right)  ,\frac{1}{\sqrt{2}}\left(  \left\vert
0010\right\rangle \pm\left\vert 1101\right\rangle \right)  ,\\
\frac{1}{\sqrt{2}}\left(  \left\vert 0100\right\rangle \pm\left\vert
1011\right\rangle \right)  ,\frac{1}{\sqrt{2}}\left(  \left\vert
0111\right\rangle \pm\left\vert 1000\right\rangle \right)
\end{array}
\right\}
\end{align}

The sixteen states span the full Hilbert space. The four sets are all mutually
orthogonal to each other. As before, we can assign a subspace to each of the
four sets spanned by the members of the respective set and therefore get the
desired decomposition. Consider now the normalized projector onto the first
subspace spanned by the set $S_{\Phi}^{+}.$%
\begin{align}
\rho^{+}  &  =\frac{1}{4}([0000+1111]+\left[  0011+1100\right] \nonumber\\
&  +\left[  0101+1010\right]  +\left[  0110+1001\right]  )
\end{align}

The permutation symmetry is obvious in the above form. Now replacing the first
two and the last two qubit states by linear combination of Bell states in
accordance to Eq (\ref{prodtobell}), one obtains,
\begin{equation}
\rho^{+}=\frac{1}{4}%
{\textstyle\sum\limits_{k=\pm}}
(\left[  \Phi^{k}\right]  \otimes\left[  \Phi^{k}\right]  +\left[  \Psi
^{k}\right]  \otimes\left[  \Psi^{k}\right]  )
\end{equation}

which we recognize as the unlockable bound entangled state presented by Smolin
\cite{smolin01}.

One can now generate three other mutually orthogonal activable bound entangled
states by applying local pauli operators on any one of the qubits of $\rho
^{+}$ using Eqs. (5, 6, 7). For instance, $\rho^{-}$ can be generated in the
following way:%
\begin{align}
\rho^{-} &  =\left(  I\otimes I\otimes I\otimes\sigma_{z}\right)  \rho
^{+}\left(  I\otimes I\otimes I\otimes\sigma_{z}\right)  \nonumber\\
&  =\frac{1}{4}([0000-1111]+\left[  0011-1100\right]  \nonumber\\
&  +\left[  0101-1010\right]  +\left[  0110-1001\right]  )\\
&  =\frac{1}{4}%
{\textstyle\sum\limits_{k,l=\pm;k\neq l}}
(\left[  \Phi^{k}\right]  \otimes\left[  \Phi^{l}\right]  +\left[  \Psi
^{k}\right]  \otimes\left[  \Psi^{l}\right]  )
\end{align}

The remaining two states $\sigma^{\pm}$ can likewise be obtained from
$\rho^{+}$ by applying the appropriate pauli operators $\sigma_{x/y}$.

\subsection{The Hilbert Space of six qubits}

Although by following our prescription, the six qubit BCABE states can be
generated from the four qubit states in a straightforward manner, here we
provide the construction from the first principles. First define the following
four sets of states:

(a) $S_{\Phi}^{+}:$ States with even number of 0 and 1 with + sign in the
superposition, like $\frac{1}{\sqrt{2}}\left(  \left\vert 000000\right\rangle
+\left\vert 111111\right\rangle \right)  ,$ $\frac{1}{\sqrt{2}}\left(
\left\vert 000011\right\rangle +\left\vert 111100\right\rangle \right)  .$

(b) $S_{\Phi}^{-}:$ States with even number of 0 and 1 with - sign in the
superposition, like $\frac{1}{\sqrt{2}}\left(  \left\vert 000000\right\rangle
-\left\vert 111111\right\rangle \right)  ,$ $\frac{1}{\sqrt{2}}\left(
\left\vert 000011\right\rangle -\left\vert 111100\right\rangle \right)  .$

(c) $S_{\Psi}^{+}:$ States with odd number of 0 and 1 with + sign in the
superposition, like $\frac{1}{\sqrt{2}}\left(  \left\vert 000001\right\rangle
+\left\vert 111110\right\rangle \right)  ,$ $\frac{1}{\sqrt{2}}\left(
\left\vert 100011\right\rangle +\left\vert 011100\right\rangle \right)  .$

(d) $S_{\Psi}^{-}:$ States with odd number of 0 and 1 with - sign in the
superposition, like $\frac{1}{\sqrt{2}}\left(  \left\vert 000001\right\rangle
-\left\vert 111110\right\rangle \right)  ,$ $\frac{1}{\sqrt{2}}\left(
\left\vert 100011\right\rangle -\left\vert 011100\right\rangle \right)  .$

Note that every group consists of sixteen members. Every group spans a
subspace that are orthogonal to each other by construction and together they
span the full Hilbert space. The decomposition is also clearly understood. Let
us now consider the unnormalized projector on the subspace spanned by the
states in the first group.%
\begin{align}
P^{+}  &  =\left[  000000+111111\right]  +\left[  000011+111100\right]
\nonumber\\
&  +\left[  000101+111010\right]  +\left[  000110+111001\right] \nonumber\\
&  +\left[  001001+110110\right]  +\left[  001010+110101\right] \nonumber\\
&  +\left[  001100+110011\right]  +\left[  001111+110000\right] \nonumber\\
&  +\left[  010001+101110\right]  +\left[  010010+101101\right] \nonumber\\
&  +\left[  010100+101011\right]  +\left[  010111+101000\right] \nonumber\\
&  +\left[  011000+100111\right]  +\left[  011011+100100\right] \nonumber\\
&  +\left[  011101+100010\right]  +\left[  011110+100001\right]
\end{align}

By construction the projector is invariant under permutation and the
normalized projector can indeed be written in a Bell-correlated form using the
Eq. (\ref{prodtobell}) :%
\begin{align}
\rho_{ABCDEF}^{+}  &  =\frac{1}{4}(\left[  \Phi^{+}\right]  _{AB}\otimes
\rho_{CDEF}^{+}+\left[  \Phi^{-}\right]  _{AB}\otimes\rho_{CDEF}%
^{-}\nonumber\\
&  +\left[  \Psi^{+}\right]  _{AB}\otimes\sigma_{CDEF}^{+}+\left[  \Psi
^{-}\right]  _{AB}\otimes\sigma_{CDEF}^{-})
\end{align}

It is now easy to construct the other three activable bound entangled states
whose unnormalized forms are projectors on the orthogonal subspaces:
\begin{align}
\rho_{ABCDEF}^{-}  &  =\frac{1}{4}(\left[  \Phi^{+}\right]  _{AB}\otimes
\rho_{CDEF}^{-}+\left[  \Phi^{-}\right]  _{AB}\otimes\rho_{CDEF}%
^{+}\nonumber\\
&  +\left[  \Psi^{+}\right]  _{AB}\otimes\sigma_{CDEF}^{-}+\left[  \Psi
^{-}\right]  _{AB}\otimes\sigma_{CDEF}^{+})\\
\sigma_{ABCDEF}^{\pm}  &  =\frac{1}{4}(\left[  \Psi^{+}\right]  _{AB}%
\otimes\rho_{CDEF}^{\pm}+\left[  \Psi^{-}\right]  _{AB}\otimes\rho_{CDEF}%
^{\mp}\nonumber\\
&  +\left[  \Phi^{+}\right]  _{AB}\otimes\sigma_{CDEF}^{\pm}+\left[  \Phi
^{-}\right]  _{AB}\otimes\sigma_{CDEF}^{\mp})
\end{align}

\section{Noisy Bell correlated activable bound entangled states}

The noisy bound entangled states are constructed by taking a convex
combination of the four BCABE states with different weights. For $2N+2$ qubits
we construct the following state :%

\begin{equation}
\rho_{2N+2}^{noisy}=\sum_{i=\pm}x_{i}\rho_{2N+2}^{i}+y_{i}\sigma_{2N+2}^{i}%
\end{equation}

where $\sum_{i=\pm}(x_{i}+y_{i})=1, \ 1\geq x_{i},y_{i}\geq0.$

Expanding the states $\rho_{2N+2}^{i},\sigma_{2N+2}^{i}$ using Eqs
(\ref{rhoplus}, \ref{rhominus}, \ref{sigmaplusminus}) one obtains,%
\begin{align}
\rho_{2N+2}^{noisy}  &  =x_{+}\frac{1}{4}\sum_{k=\pm}(\left[  \Phi^{k}\right]
\otimes\rho_{2N}^{k}+\left[  \Psi^{k}\right]  \otimes\sigma_{2N}%
^{k})\nonumber\\
&  +x_{-}\frac{1}{4}\sum_{k\neq l;k,l=\pm}(\left[  \Phi^{k}\right]
\otimes\rho_{2N}^{l}+\left[  \Psi^{k}\right]  \otimes\sigma_{2N}%
^{l})\nonumber\\
&  +y_{+}\frac{1}{4}\sum_{k=\pm}(\left[  \Psi^{k}\right]  \otimes\rho_{2N}%
^{k}+\left[  \Phi^{k}\right]  \otimes\sigma_{2N}^{k})\nonumber\\
&  +y_{-}\frac{1}{4}\sum_{k\neq l;k,l=\pm}(\left[  \Psi^{k}\right]
\otimes\rho_{2N}^{l}+\left[  \Phi^{k}\right]  \otimes\sigma_{2N}^{l})
\label{rhonoisy}%
\end{align}

which can be further expressed as
\begin{equation}
\rho_{2N+2}^{noisy}=\frac{1}{4}\sum_{k=\pm}(\Pi^{k}\otimes\rho_{2N}^{k}%
+\Gamma^{k}\otimes\sigma_{2N}^{k}) \label{rhonoisy1}%
\end{equation}

where $\Pi,\Gamma$ are two qubit Bell diagonal density matrices defined as
follows :%

\begin{align}
\Pi^{\pm}  &  =x_{+}[\Phi^{\pm}]+x_{-}[\Phi^{\mp}]+y_{+}[\Psi^{\pm}%
]+y_{-}[\Psi^{\mp}]\\
\Gamma^{\pm}  &  =x_{+}[\Psi^{\pm}]+x_{-}[\Psi^{\mp}]+y_{+}[\Phi^{\pm}%
]+y_{-}[\Phi^{\mp}]
\end{align}

The entanglement properties of such states are well known \cite{BVSW95}. Let
us note that the two qubit Werner states are special cases of the above class
of Bell diagonal states. Let $w=\max\left\{  x_{\pm},y_{\pm}\right\}  .$ Then
the states $\Pi^{\pm},\Gamma^{\pm}$ are entangled as well as distillable if
and only if $w>1/2.$ With the aid of this result we can now state the
following properties of the noisy states $\rho_{2N+2}^{noisy}:$

The state is an activable bound entangled state when $w>1/2.$ The proof is as
follows. First note that the state is invariant under interchange of parties.
This is because the state is a convex combination of the states that are
permutationally invariant. From Eq (\ref{rhonoisy}) the state is written in a
separable form across the \textit{2:2N} bipartite cut. By virtue of being
symmetric under interchange of parties, the state is separable across every
\textit{2:2N} bipartite cut. Hence the state is not distillable if all the
parties are separated from each other.

The state is entangled and distillable when $w>1/2.$ This is also clear from
Eq (\ref{rhonoisy1}). In this case, if \textit{2N} parties come together and
do collective LOCC they can distill one of the $\Pi^{\pm},\Gamma^{\pm}$ states
between the remaining two parties. However these states are distillable iff
$w>1/2.$

The states further resemble other properties of mixture of Bell states. For
instance, a mixture of two Bell state is always entangled as long as the
weights are different. Similarly one can show here by putting any two of the
coefficients $\left\{  x_{\pm},y_{\pm}\right\}  $ equal to zero that the noisy
state thus constructed is also entangled as long as the weight of the two
nonvanishing coefficients are different. However the difference is that a
mixture of Bell states is distillable while these states are only activable
and not distillable when all the parties remain separated from each other.

\section{Discussion and open problems}

To summarize we showed that the Hilbert space of even number of qubits
(greater than equal to four) can always be decomposed as a direct sum of four
orthogonal subspaces such that the normalized projectors onto the subspaces
are activable bound entangled states. The states show a surprising recursive
relation in the sense that the states belonging to \textit{2N+2} qubits are
Bell correlated to the states of \textit{2N} qubits. It is also shown that in
an activable configuration the distillable entanglement between any two qubits
is always one ebit irrespective of the total number of qubits forming the
state itself. We also studied the properties of noisy BCABE states and showed
that they are very similar to that of two qubit Bell diagonal states.

One question is immediate: Can such a decomposition be observed in the case of
odd number of qubits ? Our strategy definitely does not work in case of odd
number of qubits because the states do not have the even-even symmetry with
respect to the number of 0s or 1s in its cat/GHZ basis states. This lack of
symmetry only allows two orthogonal decomposition following our strategy but
they result into separable states.

A possible generalization of our states would be to extend to higher
dimensions. Although one can possibly do that using general pauli matrices,
the structure of such states is not immediately clear. We suspect if such a
decomposition is indeed possible then it would certainly be the number of
generators of the pauli group in dimension $d.$

As a part of future research work, one could investigate several properties of
these BCABE states. For example, one of the four qubit BCABE states (i.e.,
Smolin state) has been shown to be useful for secret key distillation
\cite{horoopenheim03}, violation of Bell inequality \cite{Horo04}, remote
information concentration \cite{muraovedral00} and superactivation of bound
entanglement \cite{sst}. We believe the results obtained in case of the
4-qubit BCABE state, could be also generalized using our states.

\vspace{1cm}

\textbf{Acknowledgement.} This work was sponsored in part by the Defense
Advanced Research Projects Agency (DARPA) project MDA972-99-1-0017, and in
part by the U. S. Army Research Office/DARPA under contract/grant number DAAD
19-00-1-0172. I.C. acknowledges CSIR, India for providing fellowship during
this work.

\end{document}